\documentclass[11pt]{article}

\usepackage{arxiv}

\usepackage[utf8]{inputenc} 
\usepackage[T1]{fontenc}    
\usepackage{url}            
\usepackage{booktabs}       
\usepackage{amsthm,amsmath,amsfonts}
\usepackage{natbib}
\usepackage{float}
\usepackage{subcaption}
\usepackage{algorithm2e}
\usepackage{xcolor}
\usepackage{multirow}
\usepackage{color,soul}
\usepackage{blkarray}
\usepackage{nicefrac}   
\usepackage[colorlinks,citecolor=blue,urlcolor=blue,filecolor=blue,backref=page]{hyperref}
\usepackage{microtype}      
\usepackage{lipsum}
\usepackage{graphicx}
\usepackage{anyfontsize}
\usepackage{afterpage}
\usepackage{adjustbox}
\usepackage{color, colortbl}
\usepackage{mathbbol}
\usepackage{mathrsfs}
\usepackage{hhline}
\usepackage{rotating}
\usepackage{comment}
\usepackage{tabularx}

\definecolor{Gray}{gray}{0.9}

\allowdisplaybreaks

\usepackage{arydshln}
\makeatletter
\renewcommand*\env@matrix[1][*\c@MaxMatrixCols c]{%
	\hskip -\arraycolsep
	\let\@ifnextchar\new@ifnextchar
	\array{#1}}
\makeatother

\RequirePackage{doi}

\renewcommand{\tilde}[1]{\widetilde{#1}}

\usepackage{bbm} 

\newcommand{\tCollapsedurl}{\if0\blind
	{
		\url{\tCollapsedurl}
	} \fi
	\if1\blind
	{ 
		\url{github.com/}\texttt{[url redacted in blinded version]}
	} \fi}

\title{Monitoring the West-Nile virus outbreaks in Italy using open-access data}

\author{
Marco Mingione\\
    \scriptsize{Dpt. of Political Sciences}\\
    \scriptsize{Roma Tre University}\\
    \scriptsize{\texttt{marco.mingione@uniroma3.it}}
\And 
    Francesco Branda\\
    \scriptsize{Dpt. DIMES}\\
    \scriptsize{University of Calabria}\\
    \scriptsize{\texttt{francesco.branda@unical.it}}
\And
    Antonello Maruotti\\
    \scriptsize{Dpt. GEPLI}\\
    \scriptsize{LUMSA}\\
    \scriptsize{\texttt{a.maruotti@lumsa.it}}
\And
    Massimo Ciccozzi\\
    \scriptsize{Unit of Medical Statistics and Molecular Epidemiology}\\
    \scriptsize{University Campus Bio-Medico of Rome}\\
    \scriptsize{\texttt{m.ciccozzi@unicampus.it}}
\And
   Sandra Mazzoli\\
    \scriptsize{STDs Centre}\\
    \scriptsize{Santa Maria Annunziata Hospital}\\
    \scriptsize{\texttt{mazzoli49@yahoo.com}}
}

\begin{document}

\RestyleAlgo{boxruled}
	
	\def\spacingset#1{\renewcommand{\baselinestretch}%
		{#1}\small\normalsize} \spacingset{1}
		
\maketitle
\begin{abstract}
This paper introduces a comprehensive and original database on West Nile virus (WNV) outbreaks that have occurred in Italy from September 2012 to November 2022. We have digitized bulletins published by the Italian National Institute of Health (ISS) to demonstrate the potential utilization of this data for the research community. Our aim is to establish a centralized open-access repository that facilitates analysis and monitoring of the disease. We have collected and curated data on the type of infected host, along with additional information whenever available, including the type of infection, age, and geographic details at different levels of spatial aggregation. By combining our data with other sources of information such as weather data, it becomes possible to assess potential relationships between WNV outbreaks and environmental factors. We strongly believe in supporting public oversight of government epidemic management, and we emphasize that open data plays a crucial role in generating reliable results by enabling greater transparency. 

	\end{abstract}
\noindent%
	
	
\section*{Background \& Summary}
West Nile Virus (WNV) belongs to the Flaviviridae family, genus Flavivirus which is a single-stranded, positive sense RNA virus \cite{rossi2010west}, and was firstly discovered in an Ugandan woman in 1937 \cite{smithburn1940neurotropic}. These viruses are named Arboviruses (Arthropod-born viruses) and are typically transmitted by bites or punctures of ticks and mosquitoes \cite{holbrook2017historical}. Before 1990, sporadic cases and mild outbreaks occurred, except in Israel and France. After 1990 several outbreaks have been observed in Algeria, Morocco, Tunisie, Italy, France, Romania, Israel and Russia, with neurological complications and deaths. In the summer of 1999, a New York cluster with its genomic sequence demonstrated the Israelian origin of the strain \cite{rossi2010west}. It is unknown how the virus crossed the Atlantic Ocean with subsequent additional spreading in Canada, the USA, Mexico, the South America Caribbean Area, Venezuela, Chile, and Argentina. South Africa and the West hemisphere are other infection cluster zones. In Italy, West Nile virus (WNV) was first detected in Toscana back in 1998 \cite{mencattelli2021west}. The regions of Emilia-Romagna and Veneto, which surround the Po River delta, were particularly affected. Since then, WNV has been detected every year in the country. To address this ongoing concern, an integrated surveillance plan for Arboviruses was initiated in the Northern Italy regions in 2008 and subsequently extended to cover the entire country \cite{calzolari2020enhanced,calzolari2022arbovirus}. \\
Various external factors may contribute to the spread of the virus, including climate change \cite{semenza2018vector}, urbanization, ease of travel, and globalization \cite{thomas2014flaviviruses}. Temperature anomalies, in particular, have been found to influence WNV transmission in Europe. They can alter the geographic range of vectors, the aerial migration routes of avian WNV hosts, and the pathogen life cycle \cite{semenza2018vector}. In the Italian climate, mosquitoes such as \textit{Culex pipiens} are considered the most likely vector for WNV transmission \cite{mancini2017specie}. The mosquito's life cycle lasts from one to four weeks and is highly sensitive to weather conditions, temperature, and fluctuations in rainfall\cite{soh2021abundance}. \\
Infections in animals, especially birds and equids, have been widely reported in the literature. Among birds, crows and other wild birds are the most commonly infected species. WNV has also been detected in rescue centers during the cold season \cite{giglia2022west}. A mosquito can acquire the virus from an infected bird and subsequently transmit it to a susceptible horse several days later. Horses typically become ill three to fourteen days after exposure to an infected mosquito (incubation period), but they cannot infect humans. Common early signs of infection in horses include twitching of the muzzle and ears, frequent chewing, aggression, and fine muscle twitching, followed by progressive lack of coordination, weakness, and listlessness. Paralysis of the limbs, seizures, disorientation, coma, or death may occur. Differential diagnosis involves considering rabies, eastern/western encephalitis, and equine protozoal myeloencephalitis (see \url{https://www.ksvhc.org/services/equine/internal-medicine/west-nile.html}).\\

\noindent All these aspects, jointly with the risk associated to contracting the virus for both humans and animals, highlight the importance of monitoring and analyzing WNV, as also demonstrated by recent research in the field \cite{tolsa2023worldwide,casades2023risk,fasano2022epidemiological,selim2021west}. \\
This paper presents an open-access spatio-temporal database, named WNVDB, which includes data on West Nile virus outbreaks occurred in the Italian territory from September 2012 up to November 2022. 
In particular, we highlight that the release of this database yields three key advantages. First, a centralized and always up-to-date repository containing all data about WNV outbreaks on the Italian territory that can be freely and easily consulted, favouring research analyses. Second, the utilization of standardized definitions and protocols (e.g., Italian administrative divisions are identified by the same codes used by the Italian National Statistical Institute (ISTAT)) allowing the users to flexibly join WNV data to other official sources of information. Third, the adherence to the FAIR (Findability, Accessibility, Interoperability, and Reusability) principles for data management and stewardship \cite{wilkinson2016fair}. 
Its scope is however wider. Having a complete open and readily available dataset about West Nile Virus in Italy can be highly useful for several reasons:
\begin{description} \item[Research and Analysis:] The dataset can serve as a valuable resource for researchers and scientists studying WNV and its impact on public health. They can use the data to analyze the patterns of WNV outbreaks, understand the transmission dynamics, identify risk factors, and develop effective prevention and control strategies. The dataset can contribute to advancing knowledge in the field and improving our understanding of the virus.

\item[Monitoring and Surveillance:] An open dataset allows for continuous monitoring and surveillance of WNV outbreaks. Public health authorities, researchers, and policymakers can regularly access the data to track the spread of the virus, identify areas of high risk, and implement timely interventions. This information can help in early detection, rapid response, and proactive measures to prevent further transmission.

\item[Collaborative Efforts:] Open data fosters collaboration among researchers, institutions, and organizations working on WNV. By sharing a comprehensive dataset, different stakeholders can collaborate, share insights, validate findings, and collectively contribute to a better understanding of the virus. This collaboration can lead to more effective strategies for disease prevention and control.

\item[Public Oversight:] Making the dataset openly available supports public oversight of government epidemic management. It enables transparency and allows citizens, journalists, and advocacy groups to examine the data, evaluate the government's response to WNV outbreaks, and hold authorities accountable. Open data promotes trust and public engagement in the decision-making processes related to public health.

\item[Improved Modeling and Predictive Capabilities:] Access to a complete and open dataset enhances the accuracy and reliability of disease modeling and predictive capabilities. Researchers can use the data to develop sophisticated models that can forecast WNV outbreaks, assess the impact of environmental factors, and guide resource allocation and preparedness efforts. Accurate models can help in early warning systems, resource planning, and targeted interventions.  
\end{description}
In summary, having a complete open and readily available dataset on WNV in Italy promotes research, monitoring, collaboration, public oversight, and the development of effective strategies to combat the virus. It empowers stakeholders with valuable information to address the challenges posed by WNV and protect public health.

The reminder of the paper is organised as follows: in the Methods Section, we describe the data sources and the protocol implemented for the construction of the database. In the Data records Section, we describe our dataset and the various associated metadata files. Finally, to show the potential of the dataset for research purposes, we discuss the results of different statistical applications in the Technical validation Section. 


\section*{Methods}
The main data sources for this study are the bulletins published in PDF format by the Italian National Institute of Health (ISS), in collaboration with Office V of the Ministry of Health's General Directorate for Preventive Healthcare and the Research Centre for Exotic Diseases (Centro studi malattie esotiche - CESME) of the Istituto Zooprofilattico Sperimentale dell'Abruzzo e del Molise "Giuseppe Caporale" (IZS Teramo). The surveillance initiative was initiated in 2012 following the enactment of \textit{DGPRE 0012922-P-12/06/2012}, which initially focused on monitoring neuro-invasive infections in humans. However, it was subsequently expanded in 2017 to include animals, specifically mosquitoes, birds, and equids. Therefore, the available data includes information on WNV infections in humans from June 2012 and in animals from August 2017.

\begin{figure}[h]
    \centering
    \includegraphics[width=.75\textwidth]{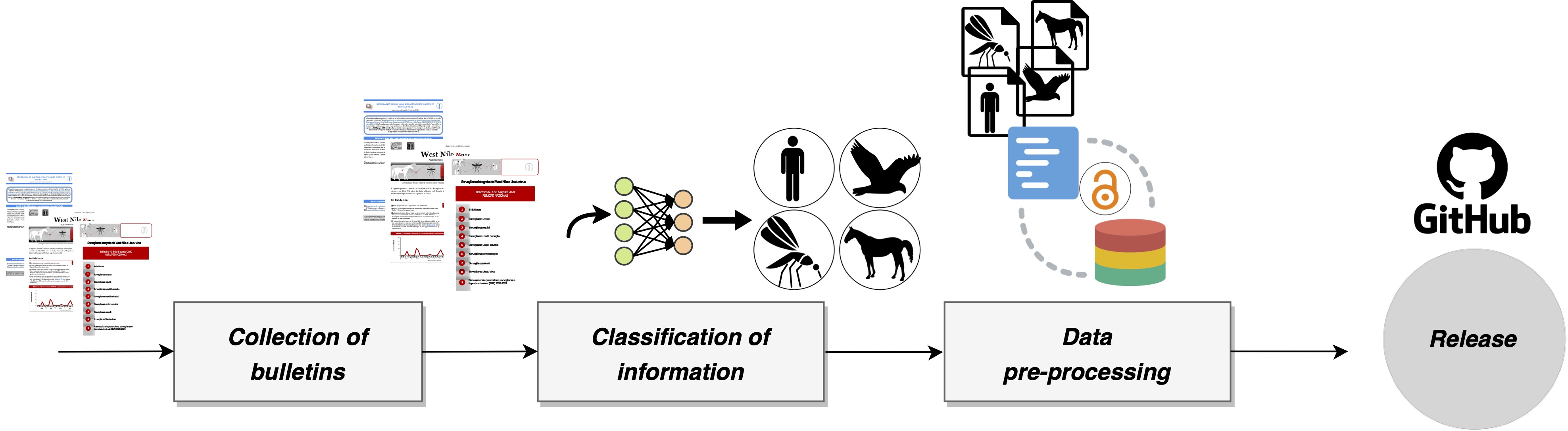}
    \caption{Schematic overview of the key passages to build the WNVDB open-access database.}
    \label{fig:methodology_schema}
\end{figure}

The data production process, covering from the digitalization to the release of WNVDB, is composed of four main steps that are summarized in Figure \ref{fig:methodology_schema}. 
In the \textit{collection of bulletins} step, we downloaded a total of 163 bulletins (up to November 2022) from the EpiCentro website, accessible at \url{https://www.epicentro.iss.it/westnile/bollettino}. After the download, in the \textit{classification of information} step, we organized these bulletins according to the corresponding surveillance category. A standard bulletin typically consists of the following sections: (i) a textual description of human cases categorized by region and infection type, (ii) a section reporting human cases by province and infection type, including a table presenting neuro-invasive cases at the provincial level categorized by age group, (iii) a section providing information on verified outbreaks in equids at the regional and provincial levels in tabular form, (iv) separate sections describing cases in target species and wild birds, and (v) a final section reporting the number of mosquitoes caught and tested positive for the virus at the regional and provincial levels. The most challenging steps were (i) and (ii), which involved extracting information from unstructured textual data. The remaining steps were also not trivial, but were facilitated by the use of an automatic tool called \textit{Tabula} (\url{https://tabula.technology/}). This tool enabled the extraction and conversion of tables from the PDF files directly into data frames. 
Afterwards, a \textit{data pre-processing} step was required to ensure coherence and consistency of our dataset. Specifically, (i) a standardized procedure was adopted to encode geographic information following the ISTAT nomenclature, also including longitude and latitude, and (ii) weekly cases have been derived by calculating the first differences of the officially reported cumulative cases (``casi\_totali'') in the bulletins. 
Here, two main issues have been addressed. First, the computation of the first differences yielded negative counts for the weekly cases (``nuovi\_casi'') of WNV: cumulative counts of epidemiological indicators are sadly known to be often affected by data quality issues mainly due to delays in reporting and/or measurement errors\cite{jona2022two}. Therefore, all negative values of ``nuovi\_casi'' have been set as Not Available (\texttt{NA}) whenever the current week`s value of ``casi\_totali'' was smaller than the value of ``casi\_totali'' in the previous week. Second, not all the information about the surveillance was available across years and for the different types of host. Indeed, missing values were sometimes present in different variables: for example, the name of the province was not always indicated for neuro-invasive human infections, but reported as ``Not indicated''; also, age-class was sometimes missing. In all these cases, we decided to leave the record in the database as higher level of aggregation (e.g. regional) could still be possible. To check the validity of our procedure, we repeatedly selected a random sample of 100 observations from the outbreak records and manually verified that the information did coincide by accessing the corresponding PDFs. In the case of a mismatch, the procedure was re-run until convergence. \\
Finally, the resulting dataset was released on a GitHub repository. The whole protocol was repeated for all bulletins, separately for each year, with the computational runtime varying from 30 minutes for early bulletins with limited data to several hours for more recent bulletins that contained additional details such as infection type and/or host information.

\section*{Data Records}
\begin{figure}[h]
    \centering
    \includegraphics[width=.9\textwidth]{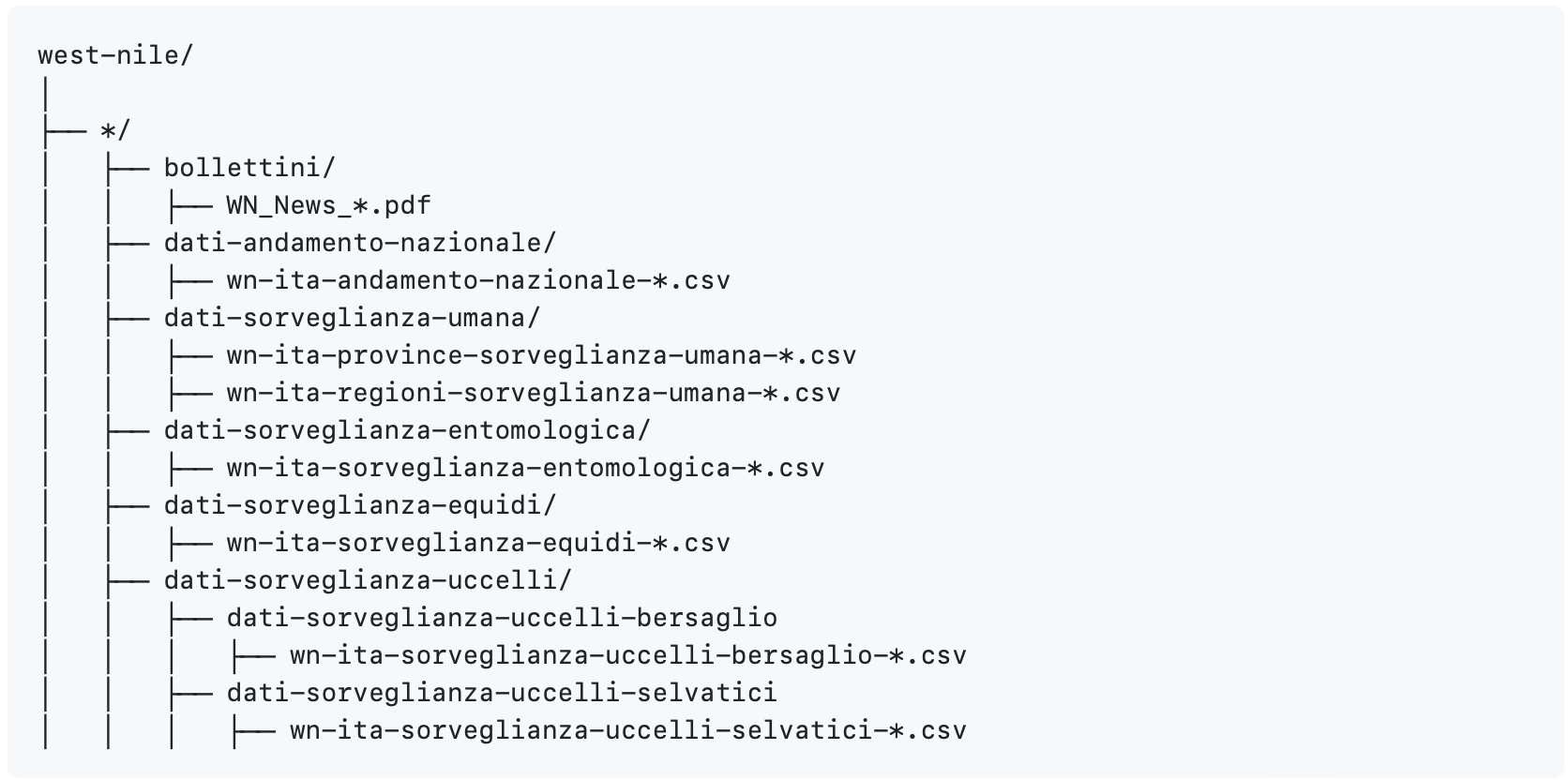}
    \caption{Schematic structure of our WNVDB.}
    \label{fig:db_schema}
\end{figure}

The database consists of one folder for each year, named from the first to the last available surveillance year. Each folder contains: (i) 
a subfolder called \texttt{bollettini} that includes all the pdf bulletins published in that year; (ii) a subfolder called \texttt{dati-andamento-nazionale} that describes the national trend of WNV cases. In addition, depending on data availability, there are up to 4 subfolders called \texttt{dati-sorveglianza-*}, where ``*'' stands for humans, equids, birds, and mosquitoes, each of which describes WNV cases per host at the regional and provincial levels. Figure \ref{fig:db_schema} shows a schematic overview of the database structure, while a more detailed description of the folders is reported below. 

\paragraph{\textbf{Bulletins (\texttt{bollettini})}}
\mbox{} \\
This folder includes the original pdf files with the data about veterinary and epidemiological surveillance of WNV as they are published by official sources. Each bulletin has been named as ``WNV\_News\_\texttt{yyyy}\_\#.pdf'', where \texttt{yyyy} is the year when the bulletin has been published and \# identifies a sequential identification number.

\paragraph{\textbf{Trend at the national level (\texttt{dati-andamento-nazionale})}}
\mbox{} \\
This folder contains data aggregated at the national level for WNV weekly and cumulative  cases. Such data are organized into .csv files, called ``wn-ita-andamento-nazionale-yyyy.csv'' (where yyyy is the year of monitoring), whose structure is reported in Table \ref{tab:table1}. In particular, each file has 5 columns and a number of rows equal to $T\times n_{\text{host}}$, where $T$ is the number of monitoring weeks and $n_{\text{host}}$ is the distinct number of host (0=humans, 1=equids, 2=target birds, 3=wild birds, 4=mosquitoes) for which data are available.

\begin{table}[ht]
    \centering
     \small
     \adjustbox{max width=.9\textwidth}{
    \begin{tabular}{|c|c|c|}
    \toprule
    \textbf{Variable} & \textbf{Description} & \textbf{Format} \\
    \midrule 
       url\_bollettino  &Link to the bulletin in pdf format	& String \\ 
        data  &Week reference date & yyyy-mm-dd \\ 
        host  &Name of host organism	& String \\ 
         nuovi\_casi  & New  cases (``casi\_totali`` current date - ``casi\_totali`` previous date)	& Numeric \\ 
           casi\_totali  & Cumulative number of cases  & Numeric \\ 

        \bottomrule
    \end{tabular}}
    \caption{Structure of the file ``wn-ita-sorveglianza-nazionale-yyyy.csv`` within the folder \texttt{dati-andamento-nazionale}.}
    \label{tab:table1}
\end{table}

\paragraph{\textbf{Human surveillance data (\texttt{dati-sorveglianza-umana})}} 
\mbox{} \\
This folder contains data about WNV infections in humans, at both the regional and the provincial level, organized into two distinct .csv files, namely ``wn-ita-regioni-sorveglianza-umana-yyyy.csv'' and ``wn-ita-province-sorveglianza-umana-yyyy.csv'' (where yyyy represents the year of monitoring). The former has a total of 9 columns and each row identifies the weekly number of cases by region and type of infection (i.e., neuroinvasive, fever, blood donor); the latter has instead a total of 13 columns and each row identifies the weekly number of WNV infections by province, age-class (i.e., $\le$ 14, 15-44, 45-64, 65-74, $\ge$ 75) and type of infection. The structure of both csv files is reported in Table \ref{tab:table2a} and Table \ref{tab:table2b}, respectively. Note that, from 2013 to 2017 only neuro-invasive cases were reported, while two more types of infection were added in the 2022 surveillance, namely  symptomatic and asymptomatic. Finally, although in principle it is possible to aggregate data to the regional level starting from the data at province level, we decided to keep both dataset separated in compliance with the bulletins' structure published by ISS.


\begin{table}[ht]
    \centering
     \small
     \adjustbox{max width=.9\textwidth}{
    \begin{tabular}{|c|c|c|}
    \toprule
    \textbf{Variable} & \textbf{Description} & \textbf{Format} \\
    \midrule 
       url\_bollettino  &Link to the bulletin in pdf format	& String \\ 
        data  &Week reference date & yyyy-mm-dd \\ codice\_regione	&Region 2-digit code &Numeric \\
        denominazione\_regione	&Region name &String \\
         lat	&Latitude of the province &Numeric \\
         long	&Longitude of the province &Numeric\\
         nuovi\_casi  & New cases (``casi\_totali`` current date - ``casi\_totali`` previous date)	& Numeric \\ 
           casi\_totali  & Cumulative number of cases  & Numeric \\ 
           tipo\_infezione  & Type of infection of reported cases & String \\ 

        \bottomrule
    \end{tabular}}
    \caption{Structure of the file ``wn-ita-regioni-sorveglianza-umana-yyyy.csv'' within the folder \texttt{dati-sorveglianza-umana}.}
    \label{tab:table2a}
\end{table}

\begin{table}[ht]
    \centering
     \small
     \adjustbox{max width=.9\textwidth}{
    \begin{tabular}{|c|c|c|}
    \toprule
    \textbf{Variable} & \textbf{Description} & \textbf{Format} \\
    \midrule 
       url\_bollettino  &Link to the bulletin in pdf format	& String \\ 
        data  &Week reference date & yyyy-mm-dd \\ codice\_regione	&Region 2-digit code &Numeric \\
        denominazione\_regione	&Region name &String \\
         codice\_provincia	&Province 3-digit code &Numeric \\
         denominazione\_provincia	&Province name &String \\
          sigla\_provincia & Province 2-letter code &String \\
         lat	&Latitude of the province &Numeric \\
         long	&Longitude of the province &Numeric\\
         eta	&Age of reported cases &String \\
         nuovi\_casi  & New cases (``casi\_totali`` current date - ``casi\_totali`` previous date)	& Numeric \\ 
           casi\_totali  & Cumulative number of cases  & Numeric \\ 
           tipo\_infezione  & Type of infection of reported cases & String \\ 

        \bottomrule
    \end{tabular}}
    \caption{Structure of the file ``wn-ita-province-sorveglianza-umana-yyyy.csv'' within the folder \texttt{dati-sorveglianza-umana}.}
    \label{tab:table2b}
\end{table}

\paragraph{\textbf{Mosquito surveillance data (\texttt{dati-sorveglianza-entomologica})}}
\mbox{} \\
This folder contains the weekly cases of WNV infections in mosquitos at the province level. It includes the file ``wn-ita-sorveglianza-entomological-yyyy.csv'' (where yyyy is the year of monitoring), the description of which is given in Table \ref{tab:table3}. 
\begin{table}[h]
    \centering
     \small
     \adjustbox{max width=.9\textwidth}{
    \begin{tabular}{|c|c|c|}
    \toprule
    \textbf{Variable} & \textbf{Description} & \textbf{Format} \\
    \midrule 
      
           url\_bollettino  &Link to the bulletin in pdf format	& String \\ 
        data  &Week reference date & yyyy-mm-dd \\ codice\_regione	&Region 2-digit code &Numeric \\
        denominazione\_regione	&Region name &String \\
         codice\_provincia	&Province 3-digit code &Numeric \\
         denominazione\_provincia	&Province name &String \\
          sigla\_provincia	&Province 2-letter code &Numeric \\
         lat	&Latitude of the province &Numeric \\
         long	&Longitude of the province &Numeric \\
         nuovi\_casi  &Total amount of new cases (``casi\_totali`` current date - ``casi\_totali`` previous date)	& Numeric \\ 
           casi\_totali  &Total number of cases  & Numeric \\

        \bottomrule
    \end{tabular}}
    \caption{Structure of the file ``wn-ita-sorveglianza-entomologica-yyyy.csv'' within  the folder \texttt{dati-sorveglianza-entomologica}.}
    \label{tab:table3}
\end{table}

\paragraph{\textbf{Equids surveillance data (\texttt{dati-sorveglianza-equidi})}}
\mbox{} \\
This folder contains weekly cases of WNV infections in equids at the province level. Differently from the surveillance of other hosts, here it is also reported the number of animals in the outbreak and the number who died. The structure of the csv file included in this folder, namely ``wn-ita-sorveglianza-equidi-yyyy.csv'' (where yyyy is the year of monitoring), is described in Table \ref{tab:table4}. 
\begin{table}[h]
    \centering
     \small
     \adjustbox{max width=.9\textwidth}{
    \begin{tabular}{|c|c|c|}
    \toprule
    \textbf{Variable} & \textbf{Description} & \textbf{Format} \\
    \midrule 
      
           url\_bollettino  &Link to the bulletin in pdf format	& String \\ 
        data  &Week reference date & yyyy-mm-dd \\ codice\_regione	&Region 2-digit code &Numeric \\
        denominazione\_regione	&Region name &String \\
         codice\_provincia	&Province 3-digit code &Numeric \\
         denominazione\_provincia	&Province name &String \\
          sigla\_provincia	&Province 2-letter code &String \\
         lat	&Latitude of the province &Numeric \\
         long	&Longitude of the province &Numeric \\
         nuovi\_casi  &Total amount of new cases (``casi\_totali`` current date - ``casi\_totali`` previous date)	& Numeric \\ 
           casi\_totali  &Total number of cases  & Numeric \\ 
            nuovi\_morti\_abbattuti  &Total amount of new deaths (``totale\_morti\_abbattuti`` current date - ``totale\_morti\_abbattuti`` previous date)  & Numeric \\ totale\_morti\_abbattuti  &Total number of deaths  & Numeric \\
           equidi\_presenti\_focolaio  &Total equidae present in the outbreak	& Numeric \\

        \bottomrule
    \end{tabular}}
    \caption{Structure of the file ``wn-ita-sorveglianza-equidi-yyyy.csv`` within the folder \texttt{dati-sorveglianza-equidi}.}
    \label{tab:table4}
\end{table}

\paragraph{\textbf{Birds surveillance data (\texttt{dati-sorveglianza-uccelli})}}
\mbox{} \\
This folder contains weekly cases of WNV infections in birds at the province level. For monitoring purposes, cases in birds are reported separately for target and wild species. The choice on the part of ISS to monitor these two types of birds is as follows: wild species can act as reservoirs and amplifiers of viral infection, while target species are sedentary so they allow viral circulation to be determined in specific areas of regions.  The folder includes two different csv files, one for each type of bird: (i) the csv file ``wn-ita-regioni-sorveglianza-uccelli-bersaglio-yyyy.csv'' (where yyyy is the year of monitoring) has data about three targeted species, i.e., \textit{Magpie} (\textit{Pica pica}), \textit{Gray crow} (\textit{Corvus corone cornix}), and \textit{Jay} (\textit{Garrulus glandarius}); (ii) the csv file  ``wn-ita-regioni-sorveglianza-uccelli-selvatici-yyyy.csv'' (where yyyy is the year of monitoring) reports WNV infections of the wild birds. Note that WNV has been found in up to 52 different wild bird species across years, but this information was not reported for the 2021 monitoring. The structure of both csv files is reported in Table \ref{tab:table5}. \\

\begin{table}[h]
    \centering
     \small
     \adjustbox{max width=.9\textwidth}{
    \begin{tabular}{|c|c|c|}
    \toprule
    \textbf{Variable} & \textbf{Description} & \textbf{Format} \\
    \midrule 
      
           url\_bollettino  &Link to the bulletin in pdf format	& String \\ 
        data  &Week reference date & yyyy-mm-dd \\ codice\_regione	&Region 2-digit code &Numeric \\
        denominazione\_regione	&Region name &String \\
         codice\_provincia	&Province 3-digit code &Numeric \\
         denominazione\_provincia	&Province name &String \\
          sigla\_provincia	&Province 2-letter code &String \\
         lat	&Latitude of the province &Numeric \\
         long	&Longitude of the province &Numeric \\
         specie	&Species name of reported cases &String \\
         nuovi\_casi  &Total amount of new cases (``casi\_totali`` current date - ``casi\_totali`` previous date)	& Numeric \\ 
           casi\_totali  &Total number of cases  & Numeric \\  
          
        \bottomrule
    \end{tabular}}
    \caption{Structure of the files ``wn-ita-sorveglianza-uccelli-bersaglio-yyyy.csv`` and ``wn-ita-sorveglianza-uccelli-selvatici-yyyy.csv`` within the folder \texttt{dati-sorveglianza-uccelli}.}
    \label{tab:table5}
\end{table}


\noindent At the time of publication, WNVDB contains more than $7,000$ records. To facilitate access to the whole available data, we created a unique dataset named ``latest-wnv.csv'' whose structure is reported in Table \ref{tab:table6}. The file has 12 number of columns and each row identifies the weekly cases of WVN infections by type of host at the province level. The idea was to provide researchers and practitioners with a compact version of the different surveillance data -- excluding specific aspects, such as age and type of infection for human surveillance or species for bird surveillance -- to allow for a quick comparison among cases by different type of host and to ease data visualization of such data to promptly identify the areas where the virus is more likely to spread in order to increase the effectiveness of prevention policies.  

\begin{table}[h]
    \centering
     \small
     \adjustbox{max width=.9\textwidth}{
    \begin{tabular}{|c|c|c|}
    \toprule
    \textbf{Variable} & \textbf{Description} & \textbf{Format} \\
    \midrule 
      
           url\_bollettino  &Link to the bulletin in pdf format	& String \\ 
        data  &Week reference date & yyyy-mm-dd \\ codice\_regione	&Region 2-digit code &Numeric \\
        denominazione\_regione	&Region name &String \\
         codice\_provincia	&Province 3-digit code &Numeric \\
         denominazione\_provincia	&Province name &String \\
          sigla\_provincia	&Province 2-letter code &String \\
         lat	&Latitude of the province &Numeric \\
         long	&Longitude of the province &Numeric \\
         nuovi\_casi  &Total amount of new cases (``casi\_totali`` current date - ``casi\_totali`` previous date)	& Numeric \\ 
           casi\_totali  &Total number of cases  & Numeric \\  
           ospite\_recettivo	&Type of infected host &String \\
          
        \bottomrule
    \end{tabular}}
    \caption{Structure of the file ``latest-wnv.csv``.}
    \label{tab:table6}
\end{table}

\section*{Technical Validation}

In this section, we show how WNVDB could be exploited to gain further understanding of West-Nile virus spread dynamics. First, we use basic summary statistics to describe the available data by type of susceptible host and provide a general picture of the virus in the Italian territory during the whole study period. Particular attention is devoted to the yearly distribution by age class (in humans) and region (in animals) of WNV weekly cases to evaluate possible trends in the spread patterns. Then, we model the yearly WNV outbreaks in 2018 and 2022 for the two most affected Italian regions, i.e. Veneto and Emilia-Romagna using the Richards’ curve \cite{richards1959flexible}. This curve, also known as Generalized Logistic Function, is very flexible and allows to characterize the main features of an epidemic outbreak (e.g. peak time, carrying capacity, etc) and to provide short-term forecasts of WNV evolution. This proposal already proved successful in modelling other epidemic outbreaks, such as the one of SARS, Dengue, Zika, Monkeypox and Ebola \cite{zhou2003severe,hsieh2004sars,hsieh2009intervention, chowell2016using,chowell2017perspectives,mingione2023short}. 

For the interested readers, we mention that WNVDB data at the province level can be easily joined with environmental factors obtained from the World Weather Online database, accessible at (\url{https://www.worldweatheronline.com/}). This demonstrates the interoperability of our dataset with other official data sources and show how to assess possible relationships between the spread dynamics and the environmental conditions. In particular, the literature suggests that the mean maximum temperature in Celsius ($^{\circ}$C), the mean total precipitation in millimeters (mm) and the mean wind speed (Km/h) affect WNV transmission in different contexts \cite{min1996progress, reisen2004west,landesman2007inter,paz2008influence,kilpatrick2008temperature,paz2013environmental}.

\begin{figure}[h]
    \centering
    \begin{subfigure}[b]{.49\textwidth}
        \includegraphics[width=.9\textwidth]{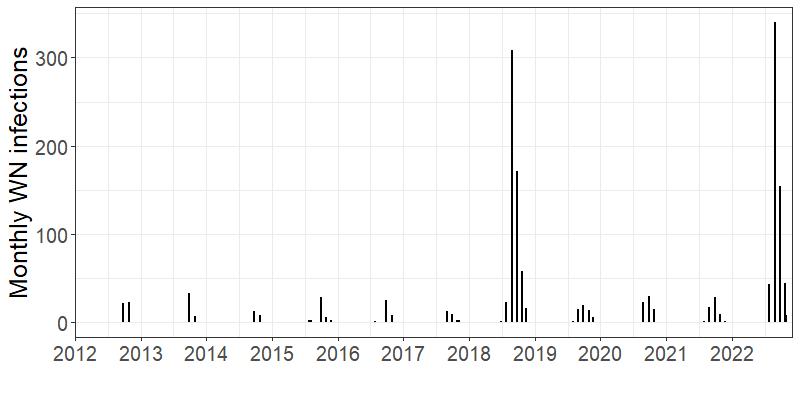}
    \caption{}
    \label{fig:hum_ts}
    \end{subfigure}
    \begin{subfigure}[b]{.49\textwidth}
        \includegraphics[width=.9\textwidth]{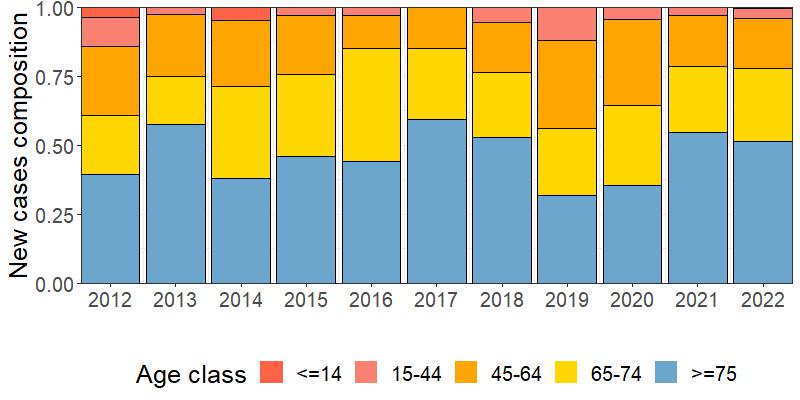}
    \caption{}
    \label{fig:humage_ts}
    \end{subfigure}
    \caption{Monthly time-series of WNV human infections (a) and yearly distribution by age-class (b) in the Italian territory.}
\end{figure}

\paragraph{Overview of WNV infections in the Italian territory over the period 2012-2022}
Since the beginning of systematic monitoring, i.e., September 27, 2012 until the most recent update on November 1, 2022, Italy counted a total of $1576$ WNV infections, with a yearly average of 143 cases. However, as it is shown by Figure~\ref{fig:hum_ts}, all outbreaks were mild except the ones of 2018 and 2022, for which a total of 581 and 599 have been recorded, respectively. Excluding these two years from the computation yields a yearly average of only 44 cases, corresponding to $< 1$ case each million residents. More than $40\%$  of human infections have been registered in Veneto (713), especially in the provinces of Padova (299) and Venezia (150), while almost $25\%$ in the region of Emilia-Romagna (407), with half of them coming from the province of Modena (118) and Bologna (97). This data highlights the importance to control other possible interrelated factors such as vector infections and environmental conditions of the infected areas. 
Figure \ref{fig:humage_ts} shows the yearly composition of human infections by age class, where the latter are: $\leq$14, 15-44, 45-64, 65-74, $\geq$75 (years). The plot depicts a clear picture where the yearly distribution by age does not substantially change. Infections in people aged $\leq$45 years are rare, while $\approx 22\%$, $\approx 27\%$ and $\approx 46\%$ of cases are recorded on average in 45-64, 65-74, $\geq$75 age classes each year. People over 75 are the ones mostly affected by symptomatic infections, as they are likely to have other age related comorbidities. Regarding the type of infections, we have that symptomatic infections (WN-s) amounts to 1417, excluding blood donors that are all asymptomatic, but including both febrile (WN-f) and neuro-invasive (WN-n) cases; in particular, we note that WN-n pathologies are prevalent almost every year -- accounting up to 87.5\% of the total WN-s in 2014 -- except for 2018 (which is also the year with the largest absolute number of WN-s).

\begin{figure}[h]
    \centering
    \begin{subfigure}[b]{.3\textwidth}
    \centering
    \includegraphics[width=.9\textwidth]{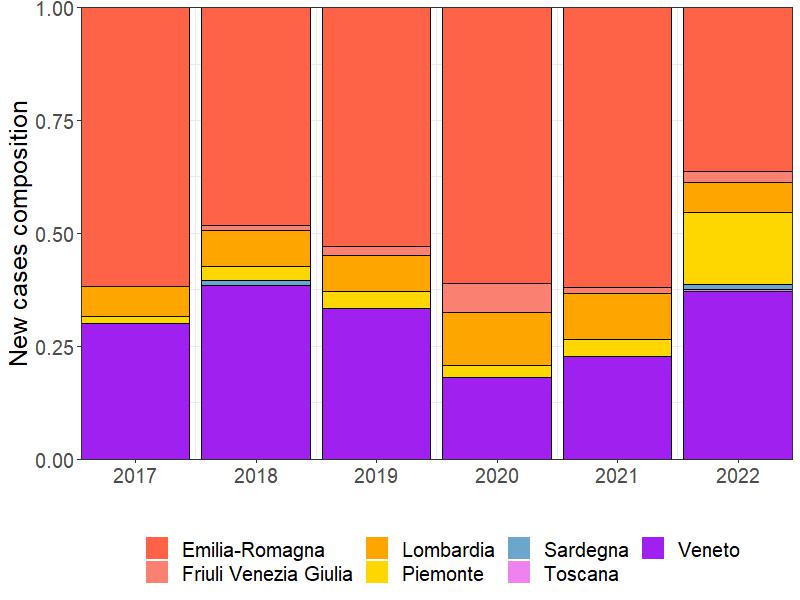}
    \caption{}
    \label{subfig:entomologica}
    \end{subfigure}
    \begin{subfigure}[b]{.3\textwidth}
    \centering
    \includegraphics[width=.9\textwidth]{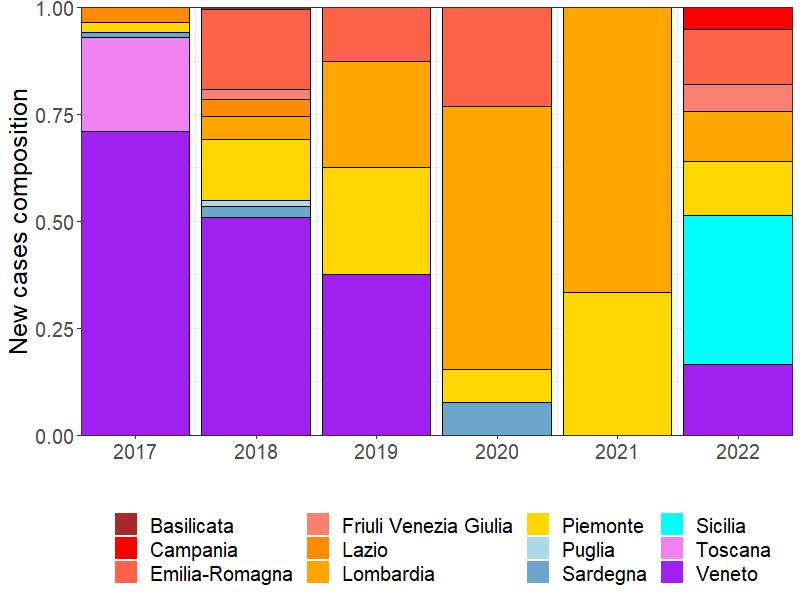}
    \caption{}
    \label{subfig:equidi}
    \end{subfigure}
    \begin{subfigure}[b]{.3\textwidth}
    \centering
    \includegraphics[width=.9\textwidth]{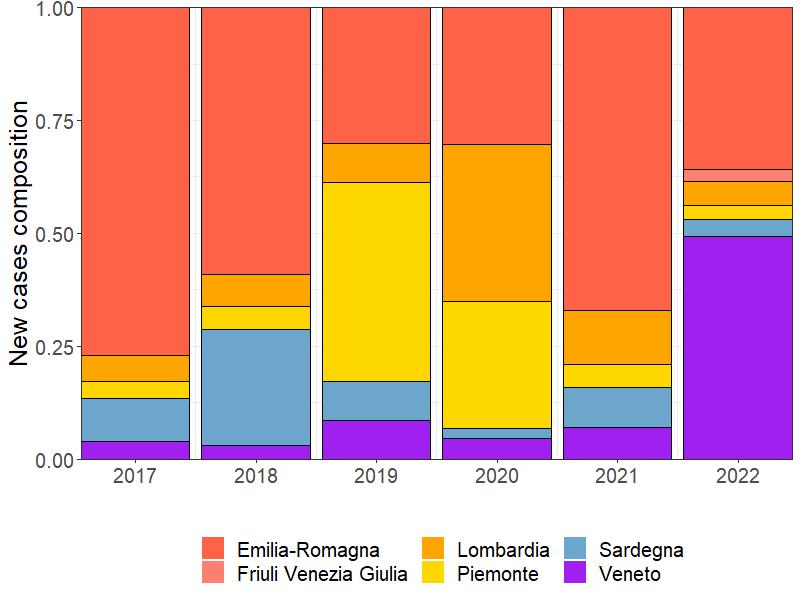}
    \caption{}
    \label{subfig:uccelli}
    \end{subfigure}
    \caption{WNV reported cases by region for each animal host: (a) mosquitoes, (b) equids, (c) birds.}
    \label{fig:compositions}
\end{figure}

Data about animal infections are only available for the last six years (2017-2022), when Italy introduced a specific prevention legislation at both the national and regional level.   
Figure~\ref{fig:compositions} reports the yearly distribution of West Nile cases by region for each animal host: (a) mosquitoes, (b) equids, (c) birds. \\
In particular, Figure~\ref{subfig:entomologica} shows that the mosquitoes infections mostly hit the three regions of Northern Italy (Emilia-Romagna, Veneto and Lombardy) with some exceptions for Friuli-Venezia Giulia in 2020 and Piemonte in 2022. Indeed, $\approx 50\%$ of WN mosquitoes infections are observed in Emilia-Romagna each year, $\approx 25\%$ in Veneto, $\approx 8\%$ in Lombardy, while the remaining $17\%$ is spread out among Friuli-Venezia Giulia, Piemonte and Sardegna (only 4 cases in 2018 and 1 case in 2022). \\
Concerning equids, the majority of the infections were detected in Veneto and Lombardia, where the horse breeding are more frequent in the territory (see Figure \ref{subfig:equidi}). Substantial decrease of WN cases in horses can be observed in Veneto after 2019, perhaps highlighting the effect of a wide campaign of prevention carried out by the region to raise awareness among horses' owners. The higher number of horses' infections in Lombardia region is instead observed in 2019 and 2020. \\
Eventually, Emilia-Romagna detected the largest number of birds infections in almost all years (see Figure \ref{subfig:uccelli}), reflecting the similar pattern of mosquitoes infections. This higher detection rate, if compared with other regions, can be due to a better application and adherence to the parameters of birds sampling in application to the regional Arbovirosis prevention law, and higher awareness of the regional governors. In 2022 the detection frequency is quite similar for Veneto and Emilia-Romagna, reflecting the higher  detection rate both in mosquitos and in human cases. 

\paragraph{\textbf{Richards' growth Generalized Linear Model}} 
Cumulative incidence of severe epidemic outbreaks typically exhibits an S-shaped trend: the onset of the outbreak anticipate an exponential growth phase whose acceleration usually softens after the implementation of one or more prevention policies (e.g. lockdown, vaccination campaign, etc.), eventually reaching an asymptote that can either be constant if the virus is eradicated or time-varying if the virus goes endemic. From another perspective, the first differences of such cumulative incidence indicators generally exhibit a \textit{bell-shaped} behaviour (\textit{wave}), still highlighting the different phases of the epidemic growth pattern. Borrowing from a well-known tool in the biological literature, we here use the Richards' curve \cite{richards1959flexible} to model weekly cases of WNV at the regional level for the two most severe outbreaks of 2018 and 2022\cite{riccardo2022rapid}. Nevertheless, we argue that this model specification can also be used to model smaller epidemic outbreaks in different regions. In particular, this curve is specified by 5 parameters able to characterize most of the features occurring in epidemiological data, such as the final epidemic outbreak size, the infection growth rate, the peak position (i.e. when the curve growth speed slows down) and the slopes of the ascending and descending phase of the outbreak \cite{tjorve2010unified}. 

For each year and for a given spatial aggregation level, let us denote the weekly time-series of cumulative West Nile cases by $\left\lbrace y^c_{t}\right\rbrace_{t=1}^T$. We model the expected value of the cumulative counts independently for each year and geographical unit by assuming that its expected value follows the \textit{extended} Richards' curve \cite{mingione2022spatio, richards1959flexible}, which can be defined as:
\begin{equation}
\label{expRich}
    \mathbb{E}[Y^c_{t}] = g^{-1}(t; \boldsymbol{\gamma}) = \lambda_{\boldsymbol{\gamma}}(t) = b \cdot t + \frac{r}{(1 + e^{h(p-t)})^{s}}, 
\end{equation}
where $\boldsymbol{\gamma}^T = [b, r, h, p, s]$ is the $5$ parameters vector including: $b\in\mathbb{R}^+$ a lower asymptote, $r>0$ the distance between the upper and the lower asymptote, $h$ the \emph{hill} (growth rate), $p\in (0,T)$ the inflation point, and $s\in\mathbb{R}$ the asymmetry parameter. Note that each cumulative counts can be defined as:
\begin{equation*}
    Y^c_t= \sum_{\tau=0}^t Y_\tau \quad\Rightarrow\quad Y_t= Y^c_t-Y^c_{t-1},\qquad t=1,\dots,T
\end{equation*}
where $Y_0=0$ without loss of generality. In other words, no matter the time scale, cumulative counts at time $t$ are obtained by summing all the \textit{new} counts from the beginning of the monitoring period up to $t$. Looking at this relationship from the opposite side, new counts at each time $t$ are the result of the difference between cumulative counts at time $t$ and $t-1$. Exploiting the linearity property of the expected value, from \eqref{expRich} we can easily derive the expected value of new counts at each time point as:
\begin{equation}
\label{eqRespFun}
    \begin{aligned}
    \mathbb{E}[Y_t] &= \mathbb{E}[Y^c_t] - \mathbb{E}[Y^c_{t-1}] =  \lambda_{\boldsymbol{\gamma}}(t) - \lambda_{\boldsymbol{\gamma}}(t-1) =  
    \tilde{\lambda}_{\boldsymbol{\gamma}}(t).
    \end{aligned}
\end{equation}
We do so to allow for the direct modelling of new counts instead of cumulative ones. It is indeed not trivial to specify a suitable probability distribution that ensures the natural constraint they must respect, i.e. $Y_t\geq Y_{\tau},\; \forall\, \tau\leq t$.
On the other hand, by assuming stochastic independence, conditionally on $\lambda_{\boldsymbol{\gamma}}(t)$, between the new counts at the time $t$ and all the previous observed values, i.e. $Y_t\,|\,\boldsymbol{\gamma}  \;\perp\; Y^c_{\tau}, \, \forall\, \tau < t$, we can easily express the likelihood of any suitable probability distribution for count data, such as the Poisson or the Negative Binomial. 
Embedding the Richards' specification in the expression of the expected value of the observed counts defines an extended GLM framework that is flexible enough to model several growth curves, such as the one of WNV cases. Parameters are estimated by numerical maximization of the log-likelihood as analytical solutions are not available in closed form. Further insights on the model implementation and estimation are provided in the seminal paper by \cite{alaimo2021nowcasting}, as well as additional details on how to compute the forecast and related uncertainty.
\begin{figure}[ht]
    \centering
    \begin{subfigure}[b]{.45\textwidth}
    \includegraphics[width=.9\textwidth]{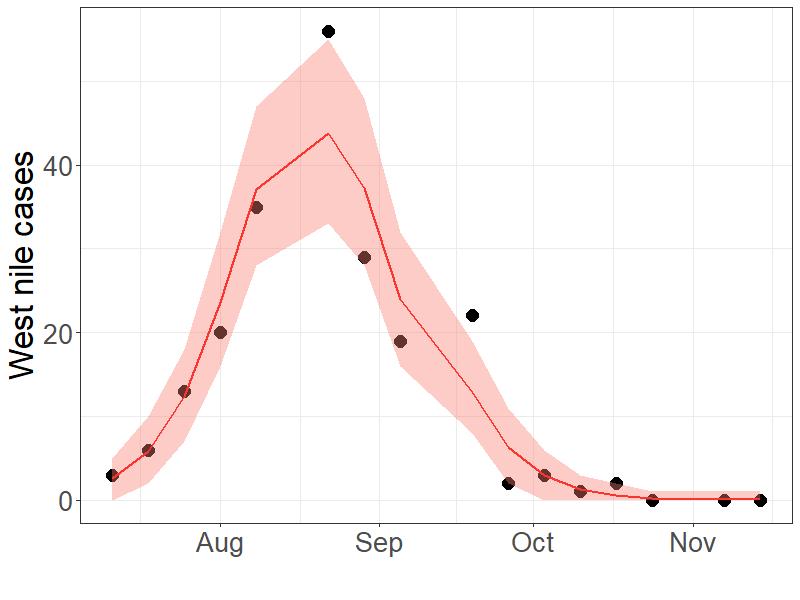}
    \caption{}
    \label{fig:estrich_emilia18}
    \end{subfigure}
    \begin{subfigure}[b]{.45\textwidth}
    \includegraphics[width=.9\textwidth]{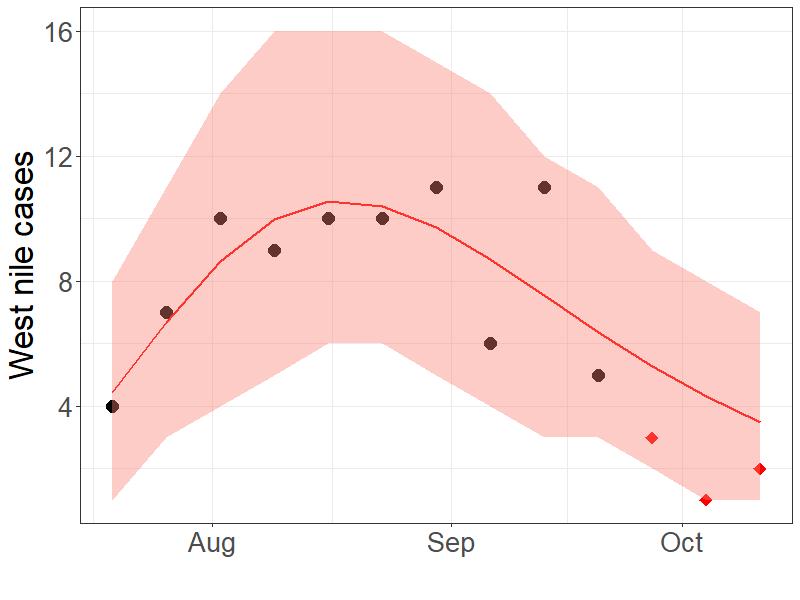}
    \caption{}
    \label{fig:estrich_emilia22}
    \end{subfigure}
    \begin{subfigure}[b]{.45\textwidth}
    \includegraphics[width=.9\textwidth]{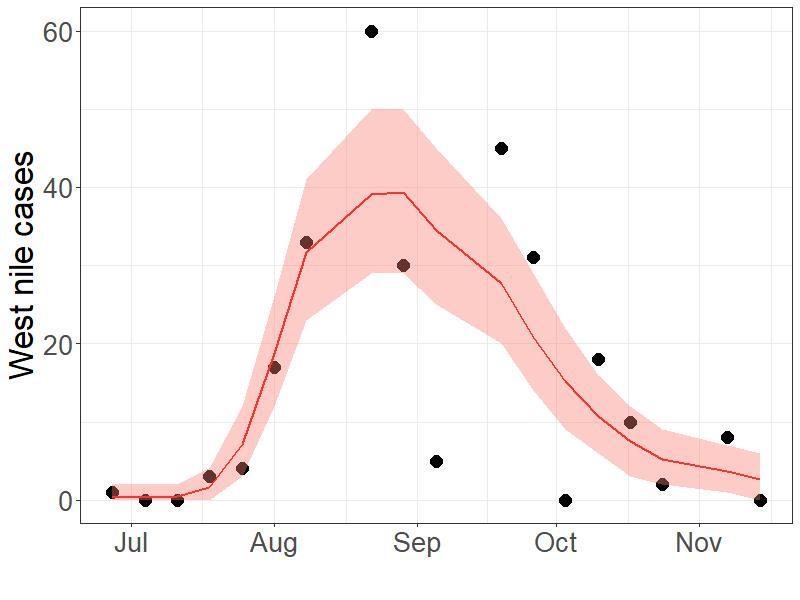}
    \caption{}
    \label{fig:estrich_Veneto18}
    \end{subfigure}
    \begin{subfigure}[b]{.45\textwidth}
    \includegraphics[width=.9\textwidth]{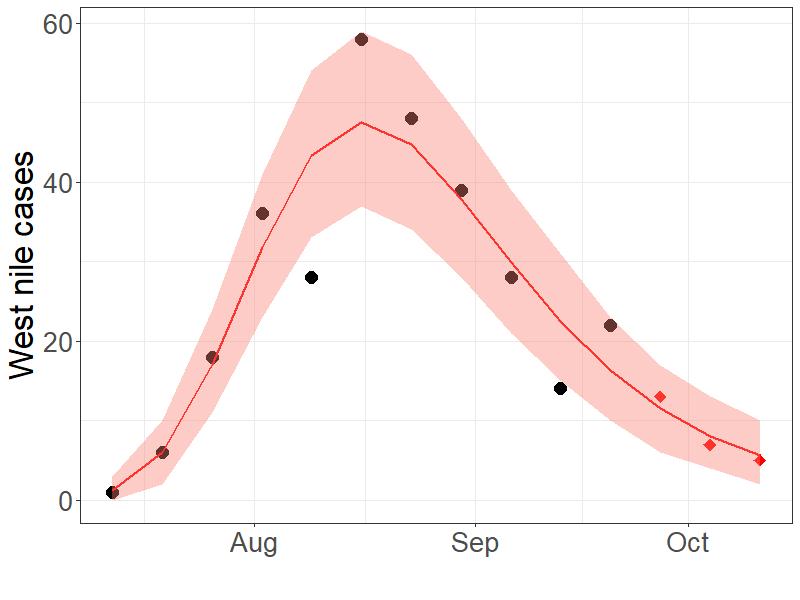}
    \caption{}
    \label{fig:estrich_Veneto22}
    \end{subfigure}
    \caption{Estimated Richards's curves of West Nile outbreak waves of (a) Emilia-Romagna in 2018, (b) Emilia-Romagna in 2022, (c) Veneto in 2018 and (d) Veneto in 2022.}
    \label{fig:estrich}
\end{figure}

Richards' parameter estimation are summarized in Table \ref{tab:richpars}.  Looking at the results of the model for the 2018 outbreak, it is possible to highlight a satisfactory goodness-of-fit for both regions, with $R^2 \approx 0.91$ and $R^2 \approx 0.61$ for Emilia Romagna and Veneto, respectively (see left panels of Figure \ref{fig:estrich}). As clear from Figure \ref{fig:estrich_Veneto18}, the management and collection of the data in Veneto is very heterogeneous, and this would affect the uncertainty surrounding our estimates. Data issues are well-known and widely discussed in the literature \cite{jona2022two}. However, the Richards' curve is able to capture the average outbreak behavior.
In detail, the estimated final epidemic size of the outbreaks in 2018 is equal to $0.4417$ (CI$_{0.95}$ = $[0.4396, 0.444]$ and $0.9104$ (CI$_{0.95}$ = $[0.9103, 0.9105]$ infections each $1000$ residents in Emilia-Romagna and Veneto, respectively, with Veneto experiencing a more at-risky situation. This is not the only difference between the two regions, as the spread of the epidemic follows a smoother behavior in Veneto than in Emilia Romagna (see Figure \ref{fig:estrich_emilia18}), with the parameter governing the infection growth rate being $\hat{h} = 0.8082$ (CI$_{0.95}$ = $[0.8081, 0.8083]$ and $\hat{h} = 0.4084$ (CI$_{0.95}$ = $[0.4084, 0.4085]$. At the same time, however, the decreasing phases is also smoother and slower in Veneto than in Emilia-Romagna, leading to a higher overall outbreak's size.
The idea that Veneto is more affected than all other Italian regions is confirmed by the analysis of the 2022 outbreak. On the other hand, the estimated final epidemic size of the more recent outbreaks in 2022 is equal to $0.30$ (CI$_{0.95}$ = $[0.019, 0.87]$ and $0.5842$ (CI$_{0.95}$ = $[0.5842, 0.5842]$ infections each $1000$ residents in Emilia-Romagna and Veneto, respectively.
The goodness-of-fit is still acceptable and consistent over outbreaks with $R^2 \approx 0.56$ and $R^2 \approx 0.84$ for Emilia-Romagna and Veneto. As clearly shown in Figure \ref{fig:estrich_emilia22}, the size of the outbreak in Emilia-Romagna is much smaller and small changes in the number of cases could lead to a more uncertain estimate of the overall outbreak size. Given the different sizes, the characteristics of the region-specific outbreaks are rather similar to those from 2018. To remark that the Richards' curve can be used not only to describe the outbreak but also to obtain short-term forecasts, we estimate the outbreak behavior leaving out the last three observations and then check if they are included in our forecast intervals. For both regions, we are able to forecast short-term events with a reasonably small uncertainty. This result can be used for future outbreaks, to monitor the evolution of the outbreak and to timely plan interventions, if required. Note that the same approach is valid for the spread of the WNV among animals too, as a further justification of the usefulness of our proposal.

\begin{table}[ht]
    \centering
      \resizebox{.95\textwidth}{!}{%
    \begin{tabular}{c|c|ccccc}
    \toprule
    \textbf{Region} & \textbf{Year} & $r$ & $h$ & $p$ & $s$ & $b$ \\
    \midrule 
        \textbf{Emilia-Romagna} & 2018 & 0.2517 (0.2505, 0.253) & 0.8082 (0.8081, 0.8083) & 5.4322 (5.4318, 5.4326) & 0.0553 (0.055, 0.0556) & 0.190 (0.1891, 0.191)  \\
        \textbf{Veneto} & 2018  & 0.0636 (0.0636, 0.0636) & 0.4084 (0.4084, 0.4085) & -16.4633 (-16.9927, -15.934) & 10 (9.7844, 10.2156) & 0.8468 (0.8467, 0.8469)  \\
        \midrule
        \textbf{Emilia-Romagna} & 2022 & 0.04 (0.005, 0.38) & 0.2537 (0.2531, 0.2543) & -20.29 (-28.69, -11.90) & 6.37 (4.25, 8.49) & 0.26 (0.014, 0.59)  \\
        \textbf{Veneto} & 2022 & 0.165 (0.165, 0.165) & 0.3909 (0.3909, 0.3909) & -3.37 (-3.38, -3.36) & 3.486 (3.484, 3.489) & 0.4192 (0.4192, 0.4192) \\
        \bottomrule
    \end{tabular}
    }
    \caption{Point estimates (95\% confidence interval) of the Richards' parameters.}
    \label{tab:richpars}
\end{table}


\section*{Usage Notes}
Our dataset is designed to promote rapid, objective, and consistent epidemiological reading of available data. This allows, for instance, to monitor the epidemiological trends of the West Nile outbreaks in Italian regions and provinces with informative graphical outputs and derive insights with predictive modeling. To facilitate other research works and ease the utilization of WNVDB, all data are stored in a GitHub repository accessible at \url{https://github.com/fbranda/west-nile} and released under a Creative Commons Attribution 4.0 International (CC BY 4.0) license, allowing users to share, copy and redistribute the material in any medium or format and to adapt, remix, transform and build upon it for any purposes. At the same link, metadata and supplementary materials are provided to better understand the dataset. As previously mentioned, this dataset was initially compiled with all data up to 2022, but will be continuously updated (and eventually adapted) as soon as new bulletins will be published. Usually, every year, the monitoring period of West-Nile cases ranges from end of May to November. Finally, following collaboration with the Agency for Digital Italy (AgID) to aggregate the database into a single portal, it can be preview and download files of interest at the link \url{https://dati.gov.it/view-dataset/dataset?id=32a9ef72-ec68-4f47-8df9-ca65c2a0a125}. Our goal is to reach as many people as possible, from the individual citizen to the professional, to evolve the current Italian open data portal into a system that provides tools and services developed and shared with the community to extend its potential.

\section*{Code availability}
All code supporting the results of this study have been developed using \texttt{R} \cite{RStatSoft} and are available on our GitHub repository at \url{https://github.com/fbranda/west-nile}, with the user instructions included in the respective ‘README.md’ files.


\begin{thebibliography}{35}
\providecommand{\natexlab}[1]{#1}
\providecommand{\url}[1]{\texttt{#1}}
\expandafter\ifx\csname urlstyle\endcsname\relax
  \providecommand{\doi}[1]{doi: #1}\else
  \providecommand{\doi}{doi: \begingroup \urlstyle{rm}\Url}\fi

\bibitem[Alaimo Di~Loro et~al.(2021)Alaimo Di~Loro, Divino, Farcomeni,
  Jona~Lasinio, Lovison, Maruotti, and Mingione]{alaimo2021nowcasting}
Pierfrancesco Alaimo Di~Loro, Fabio Divino, Alessio Farcomeni, Giovanna
  Jona~Lasinio, Gianfranco Lovison, Antonello Maruotti, and Marco Mingione.
\newblock Nowcasting covid-19 incidence indicators during the italian first
  outbreak.
\newblock \emph{Statistics in Medicine}, 40\penalty0 (16):\penalty0 3843--3864,
  2021.

\bibitem[Calzolari et~al.(2020)Calzolari, Angelini, Bolzoni, Bonilauri,
  Cagarelli, Canziani, Cereda, Cerioli, Chiari, Galletti,
  et~al.]{calzolari2020enhanced}
Mattia Calzolari, Paola Angelini, Luca Bolzoni, Paolo Bonilauri, Roberto
  Cagarelli, Sabrina Canziani, Danilo Cereda, Monica~Pierangela Cerioli, Mario
  Chiari, Giorgio Galletti, et~al.
\newblock Enhanced west nile virus circulation in the emilia-romagna and
  lombardy regions (northern italy) in 2018 detected by entomological
  surveillance.
\newblock \emph{Frontiers in Veterinary Science}, 7:\penalty0 243, 2020.

\bibitem[Calzolari et~al.(2022)Calzolari, Bonilauri, Grisendi, Dalmonte,
  Vismarra, Lelli, Chiapponi, Bellini, Lavazza, and
  Dottori]{calzolari2022arbovirus}
Mattia Calzolari, Paolo Bonilauri, Annalisa Grisendi, Gastone Dalmonte, Alice
  Vismarra, Davide Lelli, Chiara Chiapponi, Romeo Bellini, Antonio Lavazza, and
  Michele Dottori.
\newblock Arbovirus screening in mosquitoes in emilia-romagna (italy, 2021) and
  isolation of tahyna virus.
\newblock \emph{Microbiology spectrum}, 10\penalty0 (5):\penalty0 e01587--22,
  2022.

\bibitem[Casades-Mart{\'\i} et~al.(2023)Casades-Mart{\'\i},
  Holgado-Mart{\'\i}n, Aguilera-Sep{\'u}lveda, Llorente,
  P{\'e}rez-Ram{\'\i}rez, Jim{\'e}nez-Clavero, and Ruiz-Fons]{casades2023risk}
Laia Casades-Mart{\'\i}, Roc{\'\i}o Holgado-Mart{\'\i}n, Pilar
  Aguilera-Sep{\'u}lveda, Francisco Llorente, Elisa P{\'e}rez-Ram{\'\i}rez,
  Miguel~{\'A}ngel Jim{\'e}nez-Clavero, and Francisco Ruiz-Fons.
\newblock Risk factors for exposure of wild birds to west nile virus in a
  gradient of wildlife-livestock interaction.
\newblock \emph{Pathogens}, 12\penalty0 (1):\penalty0 83, 2023.

\bibitem[Chowell et~al.(2016)Chowell, Hincapie-Palacio, Ospina, Pell, Tariq,
  Dahal, Moghadas, Smirnova, Simonsen, and Viboud]{chowell2016using}
Gerardo Chowell, Doracelly Hincapie-Palacio, Juan Ospina, Bruce Pell, Amna
  Tariq, Sushma Dahal, Seyed Moghadas, Alexandra Smirnova, Lone Simonsen, and
  C{\'e}cile Viboud.
\newblock Using phenomenological models to characterize transmissibility and
  forecast patterns and final burden of zika epidemics.
\newblock \emph{PLoS currents}, 8, 2016.

\bibitem[Chowell et~al.(2017)Chowell, Viboud, Simonsen, Merler, and
  Vespignani]{chowell2017perspectives}
Gerardo Chowell, C{\'e}cile Viboud, Lone Simonsen, Stefano Merler, and
  Alessandro Vespignani.
\newblock Perspectives on model forecasts of the 2014--2015 ebola epidemic in
  west africa: lessons and the way forward.
\newblock \emph{BMC medicine}, 15\penalty0 (1):\penalty0 1--8, 2017.

\bibitem[Fasano et~al.(2022)Fasano, Riccetti, Angelou, Gomez-Ramirez,
  Ferraccioli, Kioutsioukis, and Stilianakis]{fasano2022epidemiological}
Augusto Fasano, Nicola Riccetti, Anastasia Angelou, Jaime Gomez-Ramirez,
  Federico Ferraccioli, Ioannis Kioutsioukis, and Nikolaos~I Stilianakis.
\newblock An epidemiological model for mosquito host selection and
  temperature-dependent transmission of west nile virus.
\newblock \emph{Scientific Reports}, 12\penalty0 (1):\penalty0 19946, 2022.

\bibitem[Giglia et~al.(2022)Giglia, Mencattelli, Lepri, Agliani, Gobbi,
  Gr{\"o}ne, van~den Brand, Savini, and Mandara]{giglia2022west}
Giuseppe Giglia, Giulia Mencattelli, Elvio Lepri, Gianfilippo Agliani, Marco
  Gobbi, Andrea Gr{\"o}ne, Judith~MA van~den Brand, Giovanni Savini, and
  Maria~Teresa Mandara.
\newblock West nile virus and usutu virus: A post-mortem monitoring study in
  wild birds from rescue centers, central italy.
\newblock \emph{Viruses}, 14\penalty0 (9):\penalty0 1994, 2022.

\bibitem[Holbrook(2017)]{holbrook2017historical}
Michael~R Holbrook.
\newblock Historical perspectives on flavivirus research.
\newblock \emph{Viruses}, 9\penalty0 (5):\penalty0 97, 2017.

\bibitem[Hsieh and Ma(2009)]{hsieh2009intervention}
Ying-Hen Hsieh and Stefan Ma.
\newblock Intervention measures, turning point, and reproduction number for
  dengue, singapore, 2005.
\newblock \emph{The American journal of tropical medicine and hygiene},
  80\penalty0 (1):\penalty0 66--71, 2009.

\bibitem[Hsieh et~al.(2004)Hsieh, Lee, and Chang]{hsieh2004sars}
Ying-Hen Hsieh, Jen-Yu Lee, and Hsiao-Ling Chang.
\newblock Sars epidemiology modeling.
\newblock \emph{Emerging infectious diseases}, 10\penalty0 (6):\penalty0 1165,
  2004.

\bibitem[Jona~Lasinio et~al.(2022)Jona~Lasinio, Divino, Lovison, Mingione,
  Alaimo Di~Loro, Farcomeni, and Maruotti]{jona2022two}
Giovanna Jona~Lasinio, Fabio Divino, Gianfranco Lovison, Marco Mingione,
  Pierfrancesco Alaimo Di~Loro, Alessio Farcomeni, and Antonello Maruotti.
\newblock Two years of covid-19 pandemic: The italian experience of
  statgroup-19.
\newblock \emph{Environmetrics}, 33\penalty0 (8):\penalty0 e2768, 2022.

\bibitem[Kilpatrick et~al.(2008)Kilpatrick, Meola, Moudy, and
  Kramer]{kilpatrick2008temperature}
A~Marm Kilpatrick, Mark~A Meola, Robin~M Moudy, and Laura~D Kramer.
\newblock Temperature, viral genetics, and the transmission of west nile virus
  by culex pipiens mosquitoes.
\newblock \emph{PLoS pathogens}, 4\penalty0 (6):\penalty0 e1000092, 2008.

\bibitem[Landesman et~al.(2007)Landesman, Allan, Langerhans, Knight, and
  Chase]{landesman2007inter}
William~J Landesman, Brian~F Allan, R~Brian Langerhans, Tiffany~M Knight, and
  Jonathan~M Chase.
\newblock Inter-annual associations between precipitation and human incidence
  of west nile virus in the united states.
\newblock \emph{Vector-Borne and Zoonotic Diseases}, 7\penalty0 (3):\penalty0
  337--343, 2007.

\bibitem[Mancini et~al.(2017)Mancini, Montarsi, Calzolari, Capelli, Dottori,
  Ravagnan, Lelli, Chiari, Santilli, Quaglia, et~al.]{mancini2017specie}
G~Mancini, F~Montarsi, M~Calzolari, G~Capelli, M~Dottori, S~Ravagnan, D~Lelli,
  M~Chiari, A~Santilli, M~Quaglia, et~al.
\newblock Specie di zanzare coinvolte nella circolazione dei virus della west
  nile e usutu in italia.
\newblock \emph{Vet. Ital}, 53:\penalty0 97--110, 2017.

\bibitem[Mencattelli et~al.(2021)Mencattelli, Iapaolo, Monaco, Fusco,
  de~Martinis, Portanti, Di~Gennaro, Curini, Polci, Berjaoui,
  et~al.]{mencattelli2021west}
Giulia Mencattelli, Federica Iapaolo, Federica Monaco, Giovanna Fusco, Claudio
  de~Martinis, Ottavio Portanti, Annapia Di~Gennaro, Valentina Curini, Andrea
  Polci, Shadia Berjaoui, et~al.
\newblock West nile virus lineage 1 in italy: Newly introduced or a
  re-occurrence of a previously circulating strain?
\newblock \emph{Viruses}, 14\penalty0 (1):\penalty0 64, 2021.

\bibitem[Min and Xue(1996)]{min1996progress}
Ji-Guang Min and M~Xue.
\newblock Progress in studies on the overwintering of the mosquito culex
  tritaeniorhynchus.
\newblock \emph{The Southeast Asian Journal of Tropical Medicine and Public
  Health}, 27\penalty0 (4):\penalty0 810--817, 1996.

\bibitem[Mingione et~al.(2022)Mingione, Di~Loro, Farcomeni, Divino, Lovison,
  Maruotti, and Lasinio]{mingione2022spatio}
Marco Mingione, Pierfrancesco~Alaimo Di~Loro, Alessio Farcomeni, Fabio Divino,
  Gianfranco Lovison, Antonello Maruotti, and Giovanna~Jona Lasinio.
\newblock Spatio-temporal modelling of covid-19 incident cases using
  richards’ curve: An application to the italian regions.
\newblock \emph{Spatial Statistics}, 49:\penalty0 100544, 2022.

\bibitem[Mingione et~al.(2023)Mingione, Ciccozzi, Falcone, and
  Maruotti]{mingione2023short}
Marco Mingione, Massimo Ciccozzi, Marco Falcone, and Antonello Maruotti.
\newblock Short-term forecasts of monkeypox cases in multiple countries: keep
  calm and don't panic.
\newblock \emph{Journal of Medical Virology}, 95\penalty0 (1):\penalty0 e28159,
  2023.

\bibitem[Paz and Albersheim(2008)]{paz2008influence}
Shlomit Paz and Iris Albersheim.
\newblock Influence of warming tendency on culex pipiens population abundance
  and on the probability of west nile fever outbreaks (israeli case study:
  2001--2005).
\newblock \emph{EcoHealth}, 5\penalty0 (1):\penalty0 40--48, 2008.

\bibitem[Paz and Semenza(2013)]{paz2013environmental}
Shlomit Paz and Jan~C Semenza.
\newblock Environmental drivers of west nile fever epidemiology in europe and
  western asia—a review.
\newblock \emph{International journal of environmental research and public
  health}, 10\penalty0 (8):\penalty0 3543--3562, 2013.

\bibitem[{R Core Team}(2022)]{RStatSoft}
{R Core Team}.
\newblock \emph{R: A Language and Environment for Statistical Computing}.
\newblock R Foundation for Statistical Computing, Vienna, Austria, 2022.
\newblock URL \url{https://www.R-project.org/}.

\bibitem[Reisen et~al.(2004)Reisen, Lothrop, Chiles, Madon, Cossen, Woods,
  Husted, Kramer, and Edman]{reisen2004west}
William Reisen, Hugh Lothrop, Robert Chiles, Minoo Madon, Cynthia Cossen,
  Leslie Woods, Stan Husted, Vicki Kramer, and John Edman.
\newblock West nile virus in california.
\newblock \emph{Emerging infectious diseases}, 10\penalty0 (8):\penalty0 1369,
  2004.

\bibitem[Riccardo et~al.(2022)Riccardo, Bella, Monaco, Ferraro, Petrone,
  Mateo-Urdiales, Andrianou, Del~Manso, Venturi, Fortuna,
  et~al.]{riccardo2022rapid}
Flavia Riccardo, Antonino Bella, Federica Monaco, Federica Ferraro, Daniele
  Petrone, Alberto Mateo-Urdiales, Xanthi~D Andrianou, Martina Del~Manso,
  Giulietta Venturi, Claudia Fortuna, et~al.
\newblock Rapid increase in neuroinvasive west nile virus infections in humans,
  italy, july 2022.
\newblock \emph{Eurosurveillance}, 27\penalty0 (36):\penalty0 2200653, 2022.

\bibitem[Richards(1959)]{richards1959flexible}
FJ~Richards.
\newblock A flexible growth function for empirical use.
\newblock \emph{Journal of experimental Botany}, 10\penalty0 (2):\penalty0
  290--301, 1959.

\bibitem[Rossi et~al.(2010)Rossi, Ross, and Evans]{rossi2010west}
Shannan~L Rossi, Ted~M Ross, and Jared~D Evans.
\newblock West nile virus.
\newblock \emph{Clinics in laboratory medicine}, 30\penalty0 (1):\penalty0
  47--65, 2010.

\bibitem[Selim et~al.(2021)Selim, Megahed, Kandeel, Alouffi, and
  Almutairi]{selim2021west}
Abdelfattah Selim, Ameer Megahed, Sahar Kandeel, Abdulaziz Alouffi, and
  Mashal~M Almutairi.
\newblock West nile virus seroprevalence and associated risk factors among
  horses in egypt.
\newblock \emph{Scientific Reports}, 11\penalty0 (1):\penalty0 20932, 2021.

\bibitem[Semenza and Suk(2018)]{semenza2018vector}
Jan~C Semenza and Jonathan~E Suk.
\newblock Vector-borne diseases and climate change: a european perspective.
\newblock \emph{FEMS microbiology letters}, 365\penalty0 (2):\penalty0 fnx244,
  2018.

\bibitem[Smithburn et~al.(1940)Smithburn, Hughes, Burke, Paul,
  et~al.]{smithburn1940neurotropic}
KC~Smithburn, TP~Hughes, AW~Burke, JH~Paul, et~al.
\newblock A neurotropic virus isolated from the blood of a native of uganda.
\newblock \emph{American journal of tropical medicine}, 20:\penalty0 471--2,
  1940.

\bibitem[Soh and Aik(2021)]{soh2021abundance}
Stacy Soh and Joel Aik.
\newblock The abundance of culex mosquito vectors for west nile virus and other
  flaviviruses: A time-series analysis of rainfall and temperature dependence
  in singapore.
\newblock \emph{Science of The Total Environment}, 754:\penalty0 142420, 2021.

\bibitem[Thomas et~al.(2014)Thomas, Martinez, and Endy]{thomas2014flaviviruses}
Stephen~J Thomas, Luis~J Martinez, and Timothy~P Endy.
\newblock Flaviviruses: yellow fever, japanese b, west nile, and others.
\newblock In \emph{Viral Infections of Humans}, pages 383--415. Springer, 2014.

\bibitem[Tj{\o}rve and Tj{\o}rve(2010)]{tjorve2010unified}
Even Tj{\o}rve and Kathleen~MC Tj{\o}rve.
\newblock A unified approach to the richards-model family for use in growth
  analyses: why we need only two model forms.
\newblock \emph{Journal of theoretical biology}, 267\penalty0 (3):\penalty0
  417--425, 2010.

\bibitem[Tols{\'a}-Garc{\'\i}a et~al.(2023)Tols{\'a}-Garc{\'\i}a, Wehmeyer,
  L{\"u}hken, and Roiz]{tolsa2023worldwide}
Mar{\'\i}a~Jos{\'e} Tols{\'a}-Garc{\'\i}a, Magdalena~Laura Wehmeyer, Renke
  L{\"u}hken, and David Roiz.
\newblock Worldwide transmission and infection risk of mosquito vectors of west
  nile, st. louis encephalitis, usutu and japanese encephalitis viruses: a
  systematic review.
\newblock \emph{Scientific Reports}, 13\penalty0 (1):\penalty0 308, 2023.

\bibitem[Wilkinson et~al.(2016)Wilkinson, Dumontier, Aalbersberg, Appleton,
  Axton, Baak, Blomberg, Boiten, da~Silva~Santos, Bourne,
  et~al.]{wilkinson2016fair}
Mark~D Wilkinson, Michel Dumontier, IJsbrand~Jan Aalbersberg, Gabrielle
  Appleton, Myles Axton, Arie Baak, Niklas Blomberg, Jan-Willem Boiten,
  Luiz~Bonino da~Silva~Santos, Philip~E Bourne, et~al.
\newblock The fair guiding principles for scientific data management and
  stewardship.
\newblock \emph{Scientific data}, 3\penalty0 (1):\penalty0 1--9, 2016.

\bibitem[Zhou and Yan(2003)]{zhou2003severe}
Guofa Zhou and Guiyun Yan.
\newblock Severe acute respiratory syndrome epidemic in asia.
\newblock \emph{Emerging infectious diseases}, 9\penalty0 (12):\penalty0
  1608--1610, 2003.

\end{thebibliography}



\section*{Author contributions statement}

Conceptualisation: M.M., A.M., S.M.; methodology: M.M., A.M.; formal analysis: M.M.; investigation: F.B., M.M., A.M., S.M.; data curation: F.B.; writing—original draft preparation: M.M., F.B., A.M., S.M.; supervision: A.M., S.M., M.C. All authors have read and agreed to the published version of the manuscript.

\section*{Competing interests} 
The authors declare no competing interests.

\end{document}